# The need of a "self" for self-driving cars:

# a theoretical model applying homeostasis to self-driving

Martin Schmalzried

Contact: martin.schmalzried@ucdconnect.ie

**Abstract**

This paper explores the concept of creating a "self" for self-driving cars through a homeostatic architecture designed to enhance their autonomy, safety, and efficiency. The proposed system integrates inward-focused sensors to monitor the car's internal state, such as the condition of its metal bodywork, wheels, engine, and battery, establishing a baseline homeostatic state representing optimal functionality and evaluate damage. Outward-facing sensors, like cameras and LIDAR, are then interpreted via their impact on the car's homeostatic state by quantifying deviations from homeostasis. This contrasts with the approach of trying to make cars "see" reality in a similar way to humans and identify elements in their reality in the same way as humans. Virtual environments would be leveraged to accelerate training. Additionally, cars are programmed to communicate and share experiences via blockchain technology, learning from each other's mistakes while maintaining individualized training models. A dedicated language for self-driving cars is proposed to enable nuanced interpretation and response to environmental data. This architecture allows self-driving cars to dynamically adjust their behavior based on internal and external feedback, promoting cooperation and continuous improvement. The study concludes by discussing the broader implications for AI development, potential real-world applications, and future research directions.

## 1. Introduction

7¡7¡.Background

Self-driving car technology has rapidly advanced over the past decade, transitioning from experimental prototypes to vehicles capable of navigating complex urban environments (Litman, 2020). Major automotive and technology companies are continuously refining the algorithms and sensor systems that allow these vehicles to perceive and respond to their surroundings, making them ever safer, arguably even safer than human drivers (Di Lillo et al., 2023). Despite significant progress, current self-driving systems face limitations in consistently ensuring safety and efficiency, particularly in unpredictable scenarios (Szatmáry & Lazányi, 2022).

One critical area of development is the enhancement of self-regulation and self-preservation mechanisms. Traditional self-driving car systems rely heavily on pre-programmed responses and training on the basis of data generated by external sensors to make decisions. However, these systems can struggle with real-time adaptation to dynamic environments, leading to safety risks and inefficiencies (Schwarz et al., 2021). By integrating advanced self-regulation mechanisms, modeled after biological homeostasis, self-driving cars could maintain an optimal internal state, improving their ability to navigate safely and efficiently (Yoshida et al., 2024). This approach emphasizes the importance of internal state monitoring and adaptive response strategies, setting a new direction for research and development in autonomous vehicle technology.

The concept of homeostasis, primarily borrowed from biological sciences, has been explored in various forms within the fields of artificial intelligence and robotics. However, its application specifically to self-driving cars remains relatively new. Previous research has investigated homeostatic principles in robotics to enhance autonomous behavior and resilience (Laurençon et al., 2021). These studies indicate that homeostasis can provide a framework for developing self-regulating systems capable of maintaining stability in fluctuating environments. Applying these principles to self-driving cars presents a unique opportunity to address current limitations and enhance overall system performance.

In essence, the core idea is to mimic the way a living cell navigates its external environment, striving to maintain homeostasis, as well as achieving certain goals. Just as a cell is equipped with all the necessary "sensors" to navigate through morpho-space and achieve certain morphological configurations (see morphogenesis) (Pio-Lopez et al., 2022), equipping a self-driving car with a set of inner and outer sensors, and letting it make connections between these two could help in eliciting behaviours from a self-driving car that can respond to unforeseen challenges.

This method is similar to a recent publication using a "Risk Potential Field Method" for autonomous driving (Wu et al., 2024).

7¡8¡.Problem.statement

Despite significant advancements, self-driving cars still face challenges in perception and decision-making. These systems often struggle with real-time adaptation to dynamic and unpredictable environments, which can lead to safety risks and inefficiencies. Traditional approaches rely heavily on pre-programmed responses and external sensors, but these methods are not always sufficient for handling complex, real-world scenarios.

To address these limitations, there is a growing need for self-driving cars to maintain a homeostatic state. Homeostasis, a concept borrowed from biological systems, involves maintaining internal stability amidst external changes. By adopting a homeostatic approach, self-driving cars can better manage internal states such as battery health, engine performance, and sensor functionality, thereby ensuring optimal performance and safety. This method promises to enhance the car's ability to navigate safely and efficiently, even in unpredictable conditions.

7¡9¡.Objectives

The primary objective of this research is to define a self-regulating architecture for self-driving cars that can enhance their autonomy and reliability. This involves designing a system that integrates both inward and outward-facing sensors to monitor the vehicle's internal state and external environment comprehensively. By establishing a homeostatic state, the research aims to explore how self-driving cars can maintain optimal functionality and adapt to dynamic conditions, ensuring improved safety and performance. Additionally, the study seeks to discuss the broader implications of this homeostatic approach on the behavior of autonomous vehicles, including their decision-making processes and interaction with their surroundings.

## 2. Theoretical framework

8¡7¡.Concept.of.homeostasis

Homeostasis refers to the process by which biological systems maintain a stable internal environment despite external changes. This equilibrium is crucial for the survival and optimal functioning of organisms, allowing them to respond adaptively to varying conditions. In biological systems, homeostasis involves regulatory mechanisms that balance factors such as temperature, pH, and energy levels (Michal & Klein, 2015). This concept is being explored in machine learning in order to model self-regulatory processes.

Applying these principles to artificial intelligence in self-driving cars involves creating a system where the vehicle can autonomously keep track of its internal states, such as battery health and mechanical integrity, while simultaneously adapting to external environmental changes. By mimicking the self-regulatory mechanisms found in nature - leveraging a growing field in science, biomimicry (Dicks, 2016) - autonomous vehicles could achieve a higher level of reliability and safety in diverse driving conditions, interpreting the data from sensors monitoring the outer reality via the lens of potential impact on its homeostatic state. Rather than treating all obstacles the same way, initially, and having to manually learn via human intervention which ones to ignore, self-driving cars would assess the impact of interacting with an "object" (an identified pattern) on their homeostatic state, which would make them gradually ignore obstacles like a flying plastic bag or a newspaper, based on their inner sensors checking the impact when interacting with such objects (Xiao et al., 2023). Training data could be generated via virtual environments which can more and more accurately model real-world physics (C. Hu et al., 2024) and thereby provide reliable and realistic data for determining the impact of the interaction between a virtual modelled object (such as a basketball or a newspaper) with a self-driving car, and how it affects the car's internal and structural integrity.

8¡8¡.Inward.and.outward.sensors

Inward-facing sensors are critical for monitoring the internal state of a self-driving car. These sensors track the condition of essential components such as the wheels, the metal body of the car, the engine, and battery. By continuously assessing these elements, the vehicle can detect and respond to issues like tire wear, engine performance, metal bodywork integrity, and battery health, ensuring it operates within optimal parameters and provides a "baseline" homeostatic state that it seeks to maintain. Initial calibration of these sensors and the weights determining the importance of the states of the various internal components is key, in order to elicit the appropriate behaviour from a self-driving car. For instance, protecting the integrity of its wheels and metal body frame should be very high, in order to prevent collision, while sensors turned towards internal components would mostly be used for auto-disabling (refusing to take on passengers if a problem is detected with its battery, engine etc) and auto-repairing (autonomously drive towards a garage or other maintenance facility if a problem is detected) behaviors.

Outward-facing sensors, such as cameras, LIDAR, and radar, provide the vehicle with a comprehensive view of its external environment, especially when combined (Liu et al., 2021). The visual data captured by the various sensors would be converted into several unidimensional "disks" representing alternate inner states (more on this later), as well as processed in order to recognize certain key elements such as road signs, and lane markings, which should inform their assessment for assigning risk to certain areas on these "disks" or delineate the scope of their manoeuvres based on these signs. However, the major idea is to discard the need to identify objects, focusing on converting any data into projected impact on homeostasis.

LIDAR uses laser pulses to create detailed 3D maps of the surroundings, crucial for detecting obstacles and navigating complex terrains. Radar sensors complement these capabilities by measuring the speed and distance of nearby objects, enhancing the car's ability to make safe driving decisions in real-time. Together, these sensors would provide raw data which, after being processed, would enable an AI to create a mapping of potential future projected inner states.

### 3. Training and simulation

9¡7¡.Virtual.environments

Virtual environments play a crucial role in the accelerated training of self-driving cars. These environments are designed to simulate real-world driving conditions, providing the cars with a rich dataset to learn from without the associated risks (Kaur et al., 2021).

By generating data that closely mimics real-world scenarios, these virtual environments allow for the iterative testing and refinement of a car's algorithms. The virtual training process involves exposing the cars to various driving situations, including complex intersections, pedestrian crossings, and adverse weather conditions. This helps in building a model that can potentially handle a wide range of real-world challenges.

To ensure the effectiveness of this training, the data generated in virtual environments must be highly realistic (Shen et al., 2023). This includes using high-resolution images, accurate physics simulations, and dynamic scenarios that reflect the unpredictability of real-world driving. By maintaining a close resemblance to real-world data, the virtual training process can effectively prepare self-driving cars for actual deployment on the roads.

9¡8¡.Safety.and.ethics

Safety and ethical considerations are essential in the development of self-driving cars. One key aspect is ensuring that these vehicles operate safely around living beings, and specifically, humans, which is at the heart of every concern surrounding self-driving cars (Stilgoe, 2021). To address this, artificial constraints would be introduced in the training algorithms to simulate higher damage from hitting humans. For example, if a car faces a decision between running over a pedestrian or hitting a wall, the virtual training environment would be programmed to consider hitting a pedestrian as causing near-total damage to its homeostatic state, thus ensuring that the resulting algorithm treats avoiding collision with humans as a priority.

This approach, however, introduces a potential problem: the car might become overly cautious around humans, perceiving them as significant "threats" to its homeostasis. To mitigate this, the cars could be programmed with speed thresholds. When moving below a certain speed, the cars would no longer perceive humans as threats, allowing for safer and more efficient interaction in pedestrian-heavy areas. Additionally, the cars would be trained to view humans positively; driving a human safely from one point to another contributes to maintaining the car's homeostasis by awarding it "credits" for energy replenishment and maintenance, or via some other variable which enhances the homeostatic state of the car (a type of reward function) (Silver et al., 2021).

## 4. Data conversion via machine learning

While self-driving cars employ various sensors like cameras and LIDAR to gather data about their surroundings, rather than interpreting this data through human-like perception, the car's algorithm would assign symbolic mathematical properties to the data. These properties represent potential deviations from the homeostatic state. For example, another vehicle is not seen as a car but as a potential future state that might make the self-driving car deviate from homeostasis, quantified by a numerical value (e.g., 0.3, where homeostasis is 1). This numerical approach allows the car to evaluate how different "areas" (rather than objects) in its environment might impact its homeostatic state if an accident were to occur.

One innovative aspect of this architecture is the conversion of external sensor data into a one-dimensional circular representation of reality, which could greatly optimize efficiency and data processing. This involves creating a series stacked virtual "circles" in pairs of 3 around the car, each circle having roughly 14400 data points or pixels (roughly 4 times the width of 4k resolution), where each data point on the circle represents a potential future state, ranging from completely safe (1) to extremely dangerous (0.01). Ideally, the number of disks should be around 30-42 in pairs of 3, each spaced out by about 0,5 meters, in order for the self-driving car to create a mapping of risk in space, by discrete 0,5 meter clusters. The last pair of disks would need to aggregate into a single data point, the assessed risks for all elements which are 10 meters away and farther.

These circles effectively become a dynamic "heatmap" that guides the car's navigation by indicating the safest path to maintain its homeostasis. This approach bears similarities with methods mapping out future potential risk assessments (Wu et al., 2024). However, with the existence of inner sensors monitoring the state of the self-driving car, interactions with its environment could yield to more opportunities for further training from practical experience. For instance, should the car experience a soccer ball bouncing off of its metal body, and failing to notice any impact on its inner state, it could learn to either ignore such an event/object in the future, or assign minimal risk to it prompting a minimal response when driving.

The three circle pairs would represent ground or road level danger, car level danger, and above the car (thus in total, between 90 to 120 circles in pairs of 3). Each pixel would be assigned a variable based on how quickly a detected object might pose a threat to the car's homeostatic state (rate of change of danger/safety in any given "area" in an image). Areas whose risk level change rapidly are considered closer and more dangerous, while static or slow-changing areas (in terms of risk) are deemed farther away and safer. Thus instead of creating more circles accounting for time (future projected scenarios), future risks based on rates of change are actualized in the "present moment" via multipliers or vectors of rate of change which artificially increase the perceived risk depending on rates of change of the risk. This minimizes the amount of data necessary for the self-driving car to navigate through reality, and mimics the learning process of humans which quickly learn what they should ignore, and what they should focus on, avoiding cognitive overload (Rabbitt, 2019). For instance, if an object is advancing fast and is set for a collision course with the self-driving car, the area where the collision would take place would appear as "dangerous" in the present, warranting an evasive manoeuvre in the present as well, rather than working with several layers of heatmaps accounting for time.

This conversion process involves several steps:

- Data simplification and compression: An algorithm simplifies or compresses the raw sensor data to avoid data overload (Roriz et al., 2024), focusing on essential elements that impact homeostasis based on the patterns it has identified during training. Instead of analyzing the road with high detail, it converts the raw data/information to a state ranging from safe (1) or very harmful (0.01), split into three different circles representing impact on homeostasis at the level of the road, at car height, and above the car.
- Heatmap creation: The simplified data is then used to create a heatmap around the car. Each point on this heatmap represents the likelihood of encountering a threat and the severity of the threat based on current sensor readings and predicted changes.
- Dynamic navigation map: The car uses this heatmap to navigate towards areas with higher homeostatic values while avoiding lower values. This navigation is adjusted continuously based on real-time sensor inputs.

To enhance accuracy, the car's algorithm includes a "smoothing" function that deals with the temporary persistence of perceived threats. If an area is identified as dangerous, this perception does not instantly change even if the threat disappears instantly from the external sensors' data. This lag effect ensures that temporary sensor artifacts do not lead to abrupt and unsafe manoeuvres. The car updates the status of these areas only after consistent detection over a set period, ensuring stability in decision-making. This makes the assessment of danger or risk diminish less quickly, to account for unforeseen reversal of the situation, like in the case of a pedestrian who would suddenly turn around and run in the opposite direction.

Training in virtual environments, complemented by Reinforced Learning through Human Feedback (RLHF), would ensure that these cars learn nuanced driving rules and patterns (D. Hu et al., 2024). Humans can manually label complex intersections and road segments as safe or dangerous (manually generate a heatmap conversion template on the basis of a raw image), creating a comprehensive database that the cars can use to refine their internally constructed models based on training in virtual environments. Humans can also manually compare a heatmap generated by a self-driving car to raw data, to check how the self-driving car interprets its environment, providing insight into incorrect labelling or failures to accurately identify certain threats, following with a manual correction of the heatmap allowing the car's algorithm to update its model as well.

## 5. Machine learning theoretical set-up

The machine learning approach employed to enhance the autonomy and safety of self-driving cars through a homeostatic architecture involves focusing on how the system leverages raw sensor data, processes dynamic changes over time, and integrates these insights to maintain an optimal internal state, referred to as homeostasis.

The self-driving cars would utilize a wider array of sensors to collect raw data from their environment. The reason for obtaining raw data from multiple sensors increases the chance of identifying patterns via machine learning, which could eventually lead to using only a subset of sensors, after an initial training period, examining which data was identified as relevant for making driving decisions or for generating accurate heatmaps. The raw data from these sensors would be converted into numerical values representing pixel colors (camera input), obstacle positions, and distances (LIDAR and radar), forming the foundational dataset for further analysis.

Unlike traditional methods that preprocess data to identify objects or contours, this approach allows the AI neural network to directly interact with the unprocessed numerical data. This strategy enables the AI to autonomously determine the significance of various elements in its environment and how it may impact its homeostatic state via trial and error in virtual environments, learning from its mistakes.

❶7.Dynamic.Data.Segmentation.and.Learning

The system is designed to learn from dynamic changes in the environment by analyzing overlapping one-second segments of data, updated every 50ms. This method ensures the model captures the continuous flow of environmental changes, crucial for understanding the progression leading to potential hazards or accidents.

Each 50ms, the raw sensor data is segmented into overlapping one-second windows. For instance, if the camera captures a pixel color change indicating movement or the LIDAR detects a shifting obstacle, these changes are recorded and analyzed within the context of their temporal progression. This approach allows the model to reconstruct the dynamic processes that precede accidents or other significant events, linking them to the car's homeostatic state changes based on game engine physics. The 0,05 second iteration and 1 second clustering are subject to change, and are only used as baseline initial set-up for the initial training.

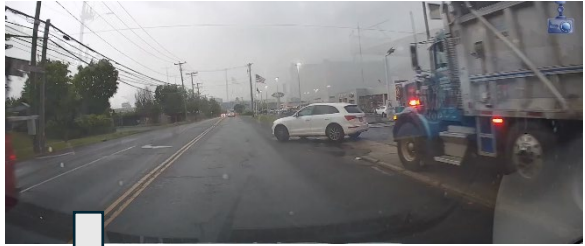

| 0.98 | 0,87 | 0,5 | 0,45 | 0,3 | 0,24 | 6?8 | 6?88 |

Rate of change vector

Assigned risk value

Identification of cluster of pixels in the range of (250,247,238)

Via comparing overlapping 1 second segments of pixel color changes, identification of displacement of cluster of pixels and increased size of the cluster (more pixels start displaying ranges (250,247,238))

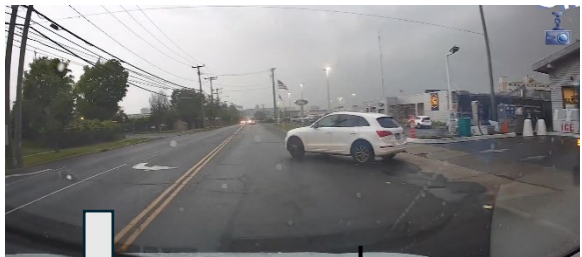

| 0.97 | 0,83 | 0,45 | 0,3 | 0,26 | 0,24 | 6?0 | 6?00 |

Through past experience with collision of data displaying similar patterns in virtual environments, assign a risk to homeostasis number along a series of 90-120 circles surrounding the car. Also assign a rate of change parameter which affects the weight of the risk value. And creates a dynamic map of risk increase/decrease approximating movement.

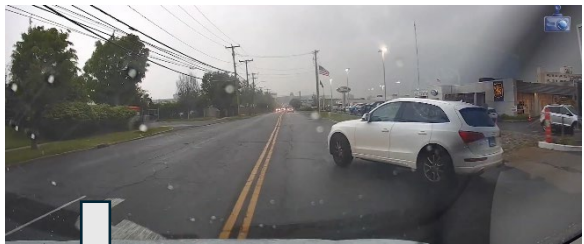

| 0.92 | 0,7 | 0,4 | 0,26 | 0,2 | 0,15 | 6?7 | 6?65 |

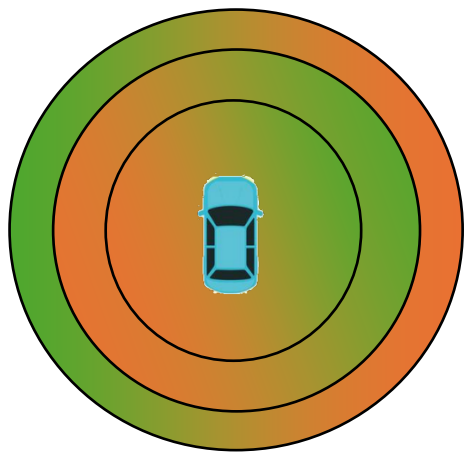

Illustration of 0,5 meter spaced circles populated by risk assessment data (impact on inner state) and rate of change of the risk.

(Green = 1, red = 0,01)

❶8.Neural.Network.Architecture

The core of the machine learning model is a neural network specifically designed to process temporal data segments and predict their impact on the car's homeostatic state. The architecture includes:

- Input layer: Receives the raw sensor data segmented into one-second windows. This layer handles multiple data streams simultaneously, such as pixel color changes from the camera, distance measurements from LIDAR, and speed data from radar.
- Convolutional layers: Extract spatial features from the raw data, capturing patterns such as object movement or environmental changes (Mersch et al., 2021). These layers are crucial for identifying potential hazards based on pixel color changes and distance variations over time.
- Recurrent layers: Implemented using Long Short-Term Memory (LSTM) units to capture temporal dependencies within the data (Kouris et al., 2020). These layers process the overlapping one-second segments, enabling the model to understand how changes in the environment evolve and match these changes to impact the car's homeostasis based on its training.
- Fully connected layers: Integrate the features extracted by convolutional and recurrent layers to make predictions about the car's future state. These layers output a numerical value representing the projected deviation from homeostasis.
- Output Layer: Produces a continuous heatmap indicating areas of potential danger (values close to 0.01) and safety (values close to 1). This heatmap guides the car's navigation by highlighting paths that maintain or restore its homeostatic state.

❶9.Training.and.Simulation

A critical aspect of training is the simulation of severe consequences for hitting humans, programming the car to prioritize human safety. This adjustment ensures that the model learns to avoid such outcomes, associating them with significant deviations from homeostasis.

One of the benefit of training in virtual environments is the ability to test the accuracy of the predictable powers of the algorithm developed during training. This can be done by matching the prediction of future impact on homeostasis from the series of circles with actual impact should the car fail to steer clear of the identified danger. For instance, in the case of a wall on the side of the road, the self-driving car's algorithm would assign the wall a certain risk metric, which represents the impact the wall would have on its homeostatic state. Simulating a crash with that wall, assuming the same conditions (speed), the risk attributed to crashing into the wall can be matched to the actual damage the car would incur (generated via game engine physics). This could be a training strategy in itself: making self-driving cars crash as many times as possible in a diverse set of scenarios and circumstances inside virtual environments in order to "calibrate" their assessment of risk and impact on their homeostatic state, and measure whether their predictive capability of perceived damage/risk matches actual damage/risk in novel untested scenarios. In other words, rather than attempting to train self-driving cars by avoiding obstacles, initially, it would be useful to forcefully let self-driving cars crash as many times as possible in order to accurately calibrate perceived risk.

## 6. Contextual metadata via a dedicated language

For the moment, language has been especially introduced to ensure self-driving cars can follow instructions from humans (Roh et al., 2020) or are able to explain their decisions in a way that is understandable by humans (Kim et al., 2021). A further step could be the introduction of a specialized "language" for self-driving cars which could offer new possibilities in how these vehicles interpret and respond to their environments.

Self-driving cars, equipped with a variety of sensors, generate extensive data about their surroundings. This data would be used, as discussed above, to inform the vehicle about potential risks and operational status, influencing the car's homeostatic state which aims to maintain optimal functionality. Traditionally, this data is processed in a binary fashion or as gradients: obstacles are either a threat or non-threat based on immediate physical parameters. However, by integrating a form of language, akin to human abstract thought, the AI responsible for self-driving in cars could categorize and respond to these obstacles in a nuanced manner, enhancing both safety and efficiency.

The "alphabet" or foundational elements of this language would likely include various operational and navigational constructs such as:

- Wheel movement;
- Speed adjustments;
- Braking;
- Acceleration;
- Directional shifts;

More abstract elements would include:

- Temporal awareness (time) or information (do X operation for Y time)
- Spatial navigation (forward and backward motions, as well as more sophisticated distance coordinates, including GPS for rerouting):
- Mathematical interpolations which serve as simplifications for movement trajectories between two points;

- If/When commands (in order to generate alternate future scenarios/responses associated to a set of actions).
- Automatic hazard notification/sharing along the lines of crowdsourcing in Waze.

Through deep learning techniques, particularly reinforcement learning within simulated environments, the car's AI would develop an internal lexicon that associates sensor inputs with specific constructs from within its "linguistic" conceptual toolbox. For instance, seeing a plastic bag flying might initially trigger a danger signal—akin to an instinctive response. However, through continuous learning, the system learns to classify this as a "non-threat," which would be expressed in its own internal language, for instance, adding metadata to the flying bag converted into a heatmap threat which says, essentially, "drive straight at constant speed", allowing for more measured responses that conserve the vehicle's resources and attention for genuine threats. In other words, it offers a second layer of sophistication to the heatmap model, enabling a self-driving car to decide on a specific course of action associated with the raw data before it is converted into a heatmap. The combination of the heatmap data, which serves as an "instinctive" layer (akin to fight or flight mechanisms in animals), with this "linguistic" or sense-making layer, could afford more nuanced reactions on the part of a self-driving car. For instance, should the data from the heatmap indicate danger, while the linguistic layer signal ignoring the threat, a third and final algorithm would reconcile or arbitrate between these seemingly contradictory instructions/data points into something akin to "caution" (for instance, proceed with driving forward but at lower speeds).

The practical application of this language involves the vehicle's AI using it to provide context to the raw sensor data. For example, both a stopped car and a fallen tree might appear as similarly risky once converted from LIDAR or camera sensors to a homeostatic risk mapping. However, the self-driving car's language would allow it to distinguish between these obstacles contextually. For instance, the stopped car might simply require to wait or circumventing the car, whereas a fallen tree could indicate more severe environmental dangers or a permanent obstacle which cannot be circumvented requiring more substantial rerouting. This could be expressed symbolically in metadata added to the heatmap in the form of "self-talk" in the language discussed above. For instance, in the case of the fallen tree, the metadata added to the heatmap would say something akin to "turn around and drive away, reroute the GPS", whereas the metadata added to the risk profile of a stopped car with a driver inside might say something like "stop and wait for X time, after X time, circumvent the obstacle".

This language could enable a richer interpretation of the car's operational environment, where data is not only collected and reacted to but also understood in a broader context. For instance, a fast-approaching object might be identified not just in terms of speed and trajectory but also in potential intent or type, allowing the vehicle to make more informed decisions about possible evasive actions.

The linguistic capabilities of self-driving cars would be designed to evolve. Through machine learning algorithms, these vehicles could continually update their internal dictionaries and syntax to refine how they interpret sensor data. This continuous learning is crucial for adapting to new scenarios and technologies, ensuring that the language grows in complexity and utility.

The linguistic nature of such a language could allow self-driving cars to react in much more sophisticated ways to obstacles besides carrying basic manoeuvres consisting in following the point of "least risk" in a mechanical way. Much like any other language, bundles of contextual information could be added to various elements at any given moment within a homeostatic risk heatmap, and also, stitched together, like human words and concepts, to form sentences and paragraphs. Various evasive manoeuvres could be added up together in a complex manner, creating relationships between them, branching out into various alternate paths or manoeuvres as events unfold.

❷7.Examples.of.proto_sentences.in.a.real.life.scenario

Imagine a self-driving car traveling on a suburban road when it encounters a situation where a large tree branch has fallen partially across the roadway just beyond a blind curve. The car's sensors detect the branch and a stopped vehicle behind it, waiting for a clear path.

The foundational metadata sentence generated to supplement the risk assessment map, using the self-driving car's language might initially read:

"Drive forward, 3.5 seconds, speed 30kph"

This initial command sets the car in motion, maintaining a cautious speed due to the suburban setting, which is crucial given the potential for unseen obstacles just beyond a blind curve.

Branching.Scenarios¿

- If the branch is detected with no other obstacles:

"Reduce speed to 10kph, drive forward 5 meters, stop, send alert for hazard"

This sequence commands the car to slow down considerably, approach cautiously, come to a stop a safe distance from the branch, and alert nearby vehicles or a central traffic management system about the hazard.

- If the branch is detected along with a stopped car behind it:

"Slow to 5kph, drive forward 4 meters, stop, send alert of hazard"

Here, the car further reduces its speed due to the increased complexity of the scene and closes in slightly to better assess whether the stopped car is about to maneuver around the obstacle or remain stationary.

- If the stopped car begins to maneuver around the branch:

"Hold position, wait 10 seconds "

This command ensures that the self-driving car holds its position to avoid any potential collision with the maneuvering car and reassesses the situation after giving the other vehicle some time to clear the path.

- If the path clears:

"Speed to 30kph, drive forward"

Once the path is deemed clear, the car resumes driving forward, but at lower speeds for a situation reversal.

- If the branch remains and the stopped car has not moved:

"Reverse manoeuvre, reroute via alternate road"

This scenario calls for a retreat and rerouting if the path remains blocked and there are no indications of immediate clearance after a set time..

Each of these proto-sentences utilizes the "alphabet" of self-driving car language, comprising operational manoeuvres like speed adjustments and directional shifts, integrated with higher-level strategic decisions like stopping, waiting, or rerouting. The AI's ability to string these actions into coherent "sentences" using programming languages like Python could enable it to navigate complex real-world situations much like a human driver using reasoning and adaptation, but based on a quantifiable assessment of risks and environmental data based on learning from within virtual environments.

The "mastery" over the grammar, syntax and utility of the language in diverse virtual scenarios would be fine tuned thanks to feedback from the materialisation of a risk (failure to avoid an obstacle, or failure to maintain its homeostatic state). Via thousands of trial and error simulations, an algorithm could learn to leverage such elements of "car language" to refine its reactions in more granular ways, supplementing the data gathered from the risk mapping exercise.

Some possible initial parameters, or grammar to enable the deep learning algorithm to train in virtual simulations, would be to let the self-driving car associate metadata to heatmaps it generates every 50ms, projecting itself about 10 seconds into the future (programming its intended actions for the next 10 seconds). For example, raw images from a camera show an animal on the side of the road. The image is converted into a

"risk to homeostasis" heatmap, with a simultaneous generation of metadata issuing commands about 10 seconds into the future, associated to alternate predictive scenarios of the heatmaps' evolution. For instance, the self-driving car could generate a scenario whereby the animal runs away from the road, and the associated instructions would read "maintain speed, 10 sec". Or the self-driving car could generate another scenario whereby the animal runs in the middle of the road, which would include alternate metadata such as "brake steadily until speed is 0 over 3 seconds". Every 50ms, the raw data generated by sensors either strengthens or weakens the likelihood of these alternate scenarios, which the self-driving car would start implementing, leading to a probabilistic convergence towards a future scenario which best matches the actual evolution of the situation. This could be visually represented as alternate "timelines" branching out in multiple scenarios generated every 50ms, some being reinforced by the raw image data, some being weakened, and a "main" scenario which represents the actual behaviour of the self-driving car, based on interpolating through these alternate scenarios every 50ms depending on the evolution of the situation as picked up by its sensors and converted into a risk heatmap.

A lightning bolt could be used to illustrate this process: much like a lightning bolt that seeks the path of least resistance to connect the clouds to earth, the self-driving car, guided by its language and continuous sensor data, navigates its environment to connect its present state to an optimal future state (its point of arrival or intended destination). This connection is achieved through dynamic adjustments and probabilistic convergence, ensuring a safe and efficient journey.

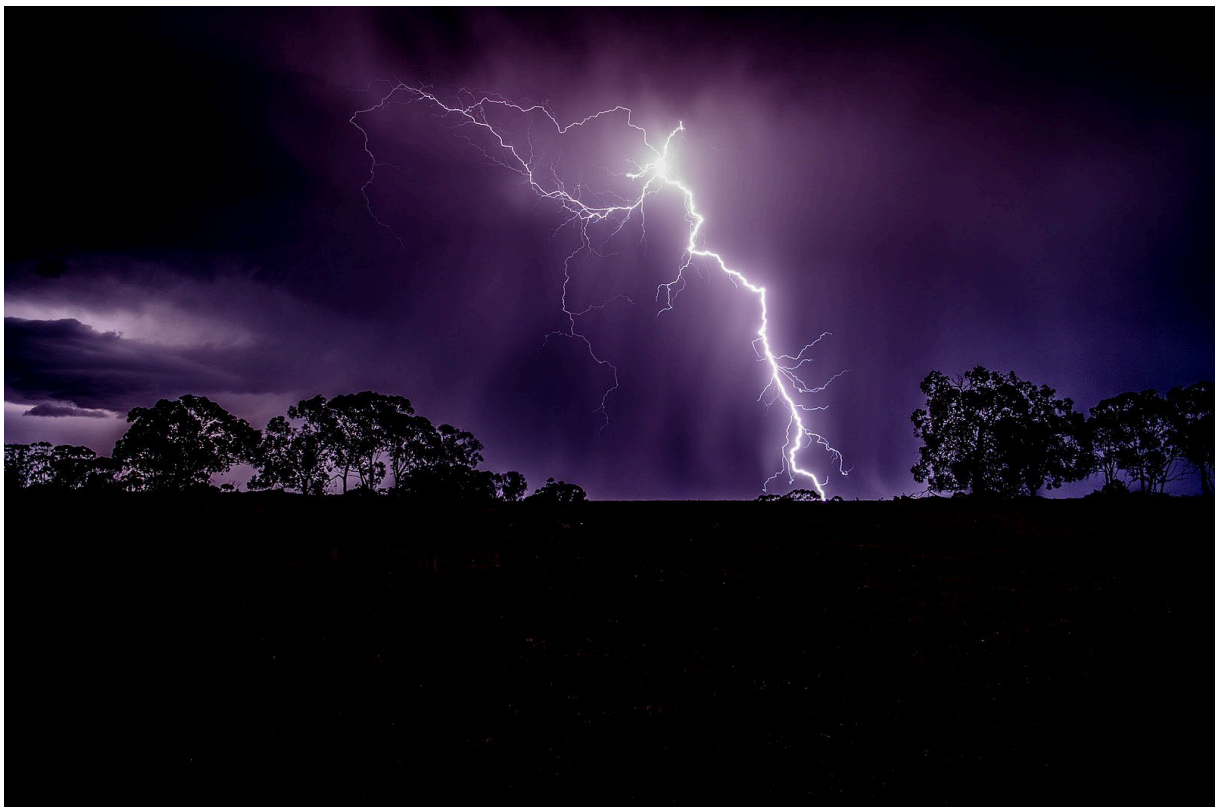

## 7. Behavioral Adaptations

**❸7¡.Human.Interaction**

One of the critical aspects of developing autonomous vehicles is ensuring that they can interact safely and efficiently with humans. This involves programming self-driving cars to recognize and respond to human presence in a manner that prioritizes safety and enhances the vehicle's operational effectiveness.

To achieve this, self-driving cars are equipped with sophisticated algorithms that allow them to identify humans using data from their outward-facing sensors, such as cameras and LIDAR, leveraging existing work done in facial recognition, while being mindful of the privacy issues involved (Eastman, 2022). To manage these interactions effectively, it is essential to establish thresholds for speed and proximity. At higher speeds, the car is programmed to be more cautious, maintaining a greater distance from pedestrians and other vulnerable road users. Conversely, at lower speeds, the car can afford to be closer to humans without posing a significant risk. Given that the algorithm would be tweaked to paint humans as the "most lethal" to the homeostatic state of self-driving cars, their natural behaviour would be "fear" of humans (or in other words, seeking to maximize the distance between them and humans), which defies the purpose of self-driving cars (driving humans around).

Thus a more nuanced programming should be introduced so that human presence, within certain parameters, is viewed as a positive factor. By associating the transportation of humans with positive reinforcement, such as a reward function which can be converted into actions which positively affect its homeostatic state such as maintenance or battery recharging, as well as creating a threshold below which humans cease to represent a superior "danger" to the car (when it drives at very low speeds), the self-driving car would be encouraged to serve its passengers efficiently while maintaining safety.

**❸8¡.Cooperation.and.Communication**

Another vital component of behavioral adaptation in self-driving cars is the ability to communicate and cooperate with other autonomous vehicles on the road. Effective inter-car communication can significantly enhance the overall safety and efficiency of the traffic system.

Self-driving cars should utilize Vehicle-to-Vehicle (V2V) communication technologies based on local mesh networks, bypassing centralized communication networks (Ulhe et al., 2020), to share information about their surroundings (relevant dangers), current state and intentions. This information includes speed, intended pathway (sharing a vector which represents their intended direction), location, and any detected areas within it's generated heatmap which represent significant danger. Sharing raw data from their sensors is challenging and impractical for strategic V2V communication(Ngo et al., 2023). By continuously exchanging data in the form of relevant risks in their own heatmap, reinterpreted based on the differential position/location, the cars can anticipate each other's actions and make coordinated decisions that reduce the risk of accidents and improve traffic flow.

To manage the quality and utility of the shared information, a reward/penalty system should be implemented. This system assesses the usefulness of the data provided by each car in maintaining homeostasis and achieving other operational goals, such as minimizing travel time and energy consumption. For instance, if a car shares information about a road hazard that is validated by another car as information that was not registered or noticed by its own model, but only appeared later (once the situation evolved) and thus helped anticipate or prevent an accident, it receives a reward. On the other hand, if a car disseminates irrelevant or incorrect data, it incurs a penalty, discouraging the spread of unhelpful information.

This reward/penalty function ensures that the communication network remains reliable via encouraging the car's algorithms to be selective and accurate in the data they share, promoting a high standard of information exchange.

In addition to V2V communication, self-driving cars should also be capable of learning from shared experiences, introducing some form of memory of key events/data (Chen et al., 2017). When a car encounters an unusual or challenging situation, such as having its inner homeostatic state affected without anticipating it, it records the raw data from the event, the heatmap generated from the raw data, the series of instructions (metadata) that the car added to the heatmap, and the outcome (impact on its homeostatic state), which is then shared on a blockchain accessible to other autonomous vehicles. Human agents would review and validate these "uploads" in order to add a manually generated heatmap fitting the situation, and provide a series of possible alternate manoeuvres expressed in car language. Each car can evaluate the relevance of this shared experience based on its own internal memory and context, ensuring that the learning process is decentralized and tailored to individual operational models. Instead of creating a single centralized model spread uniformly across the self-driving car fleet, the existence of local custom models which self-actualize based on select shared data should be encouraged to promote dynamic evolutionary mechanics, akin to "natural selection" based on such parameters as the ability to arbitrate effectively between all the goals: maintain homeostasis, get to set destination, use least amount of energy, minimize time of the trip, respect speed limitations.

Positive reinforcement should also be mirrored by learning from negative incidences. Memory and learning are crucial components for the advancement of self-driving car technology, enabling vehicles to continuously improve their decision-making processes through the sharing of experiences, particularly accident data. This would be facilitated through the use of blockchain technology, which provides a secure and decentralized method for recording and disseminating this information. Such memories can be integrated during maintenance, and their relevance tested via simulations to ensure that models are not adversely affected.

To account for the size of the data to be uploaded, decentralized storage solutions such as Sia (Khalid et al., 2023) could be leveraged for certain data (such as the raw footage of 10-20 seconds providing the entire context for the accident).

Each self-driving car would utilize its own model to interpret the shared experiences made available via this public blockchain. The car's internal algorithms analyze the new data in the context of its own algorithm, sensor configurations, and operational history. This personalized assessment helps the car determine the applicability of the shared information to its unique circumstances, which further stimulates cross-fertilization and evolution of these models. For instance, a car that frequently operates in urban environments may prioritize data from similar settings over experiences from rural areas. This localized relevance filtering ensures that each car can effectively integrate useful lessons while ignoring less pertinent information.

Individualized training and validation can help ensure that self-driving cars are better equipped to handle specific challenges of their typical operating environments and encourages a diversity of responses and adaptations across the fleet, reducing the likelihood of systemic failures that could occur if all cars were programmed identically. However, self-driving cars should be periodically tested on virtual simulations (for instance, during maintenance) to detect any anomaly in their individual algorithms/models. Humans would be responsible for designing more and more sophisticated virtual environments and simulations to track and "debug" the reliability of the individual algorithms of each self-driving car, with a benchmark of acceptable errors per simulation cycle (for instance, simulating driving for 100 hours in a virtual city).

### 9. Human oversight and human safety

Human oversight is also crucial in validating shared experiences to prevent erroneous or harmful behavior patterns from spreading. For example, if a car is attacked by vandals, the incident needs careful evaluation to ensure that the car's response does not lead to an unjustified generalization that all humans are threats. On the other hand, encounters of self-driving cars with rare scenarios such as a rugby ball bouncing off a car, leaving it unscathed, is a memory worth sharing, as it would prevent a self-driving car from reacting via a dangerous avoidance manoeuvre which could endanger the safety of the passenger.

In this regard, a self-driving car's algorithm should also account for two different scenarios: driving without any passengers, driving with passengers, and tweak its driving behaviour accordingly. In case a self-driving car is carrying passengers, then its risk heatmap should be modulated based on threats to the safety of passengers. In virtual settings, data used from car crash dummies could be leveraged to model the impact on human health and safety in case of an accident, and the homeostasis state of the car would be modulated to account for a homeostasis indicator for the humans it carries (Jaśkiewicz et al., 2021). For instance, if a self-driving car has the choice between avoiding a big rock which would smash its windshield with a very rough manoeuvre or let the rock smash the windshield while decelerating progressively, it would choose the later in order to protect the safety and health of humans inside of the car.

The operationalization of this system would be challenging, as the trade-off between maintaining homeostasis for the self-driving car while simultaneously safeguard the health and safety of humans can be tricky, and could result in unforeseen behaviours where a self-driving car could make unpredictable decisions based on certain dilemma (for instance, choosing between running over a pedestrian and safeguarding the health of humans inside of the car, or the reverse) (Servin et al., 2023). In this sense, manually creating specific virtual preprogrammed scenarios whereby a self-driving car is driving at very high speeds while carrying passengers and a pedestrian runs in front of the vehicle, could help in providing manually validated training data, where humans would directly survey the self-driving car's reaction, the heatmap generated from the experiment, and manually correct these two pieces of data with a more appropriate heatmap and more appropriate associated reactions (avoiding the pedestrian, hitting the guardrails etc).

Finally, reinforced learning techniques via either manual human labelling should only be applied to situations whereby alternate courses of action could have been taken to avoid a collision. Sharing and validating crash scenarios where self-driving cars had no alternate avoidance actions available could lead to the inadvertent interpretation that these scenarios are explicitly recommended or "allowed" inside the model.

### 30.Motivation

Given that the car's main objective is to maintain homeostasis, this could result in cars remaining static. To solve such an issue, homeostasis should be broken down into hierarchies, whereby the integrity of the outer metal body shell and wheels is deemed highest (to avoid collision), integrity of the motor and batteries is ranked lower. To incentivise movement, whenever a destination is encoded into the self-driving car, the perception of its batteries' homeostatic state is artificially reduced by the energy it would cost the vehicle to make the trip. For instance, if the trip is projected to consume 30% of the car's batteries, then the state of the battery would be virtually reduced by 30% + a premium (to ensure the car has sufficient motivation to reach its destination).

However, the self-driving car should never consider that recharging its batteries (restoring the homeostatic state) is more important than maintaining the integrity of its metal bodywork, otherwise it may lead to behaviors whereby a self-driving car with very low battery levels might collide with an obstacle in order to get to a charging station faster, considering that the "trade-off" is worth it.

Once the car arrives at a set destination, the car would receive "charging credits" via a reward function which would compensate for the loss in battery life which would be used to replenish its battery levels once it reaches a certain threshold. This system could be further tweaked depending on the car's behavior, even allowing the car to gain sufficient autonomy to "take care" of itself (decide when to go back to the charging station to convert the credits it has accumulated in order to recharge its batteries fully, depending on its distance from charging stations).

This system would mimic navigation and motivation for biological creatures, who navigate through their environments seeking food/energy to restore or maintain their homeostatic state. In this sense, a destination on a map would be perceived, symbolically, by a self-driving car's algorithm, as energy that it needs to maintain homeostasis, or in other words, giving the car a "telos" by creating an attractor, or an end state that it seeks to achieve, given certain constraints (Levin, 2022).

## 8. Discussion

### 7. Implications for AI Development

The integration of homeostatic principles into self-driving car systems could represent a significant advancement in the field of artificial intelligence and autonomous systems. This approach could impact the future design of autonomous systems by introducing a framework where vehicles can maintain an optimal state of functionality through continuous self-regulation, mimicking the mechanism of homeostasis in biology, via sensors monitoring the internal state of autonomous systems. By monitoring and responding to both internal states and external environments converted into potential future alternate internal states, these systems can make more informed decisions that enhance safety, efficiency, and overall performance.

One of the critical implications for AI development is the shift towards more adaptive and resilient systems. Traditional autonomous systems often rely on pre-programmed responses via training on human labelled data and static models, which can limit their ability to handle unexpected scenarios. The homeostatic approach, however, allows for dynamic adjustments based on real-time data, potentially fostering a higher degree of autonomy. This adaptability is crucial for the safe deployment of self-driving cars in diverse and unpredictable real-world conditions.

Ethical considerations also play a vital role in the development and deployment of these systems. Ensuring that autonomous vehicles operate safely around humans is essential. The introduction of artificial constraints to simulate higher damage from hitting humans could address the ethical imperative of prioritizing human safety. However, this also raises concerns about potential over-cautious behavior, where cars might avoid humans excessively, leading to inefficiencies or even operational anomalies. Balancing these safety protocols with practical performance is a delicate task that requires ongoing refinement and ethical oversight.

Adoption of homeostatic self-driving cars could potentially reduce traffic accidents, lower emissions through optimized energy consumption, and enhance mobility for populations that are currently underserved by traditional transportation options.

## 8. Limitations and Future Work

While the proposed homeostatic system for self-driving cars could offer some advantages, it is not without its limitations. One of the primary challenges is the complexity of accurately modeling the homeostatic state and ensuring that the car's sensors and algorithms can reliably maintain this state and especially accurately convert raw data from sensors to homeostatic "risk" in all driving conditions. The variability of real-world environments presents significant challenges, as the car must continuously adapt to changing conditions such as weather, road quality, and unpredictable human behaviors.

Another limitation is the reliance on virtual environments for initial training. Although these environments can closely mimic real-world conditions, they cannot capture every possible scenario a car might encounter. This gap between simulated training and real-world application means that continuous learning and adaptation are necessary once the cars are deployed. Ensuring that the transition from virtual to real-world driving is seamless and effective remains an ongoing challenge.

Future research should focus on improving the accuracy and reliability of the homeostatic models used in these systems. This includes refining the algorithms that convert sensor data into actionable insights and enhancing the car's ability to predict and respond to dynamic changes in its environment. Additionally, developing more sophisticated virtual training environments that can encompass a broader range of scenarios will be crucial for preparing self-driving cars for real-world conditions.

Another area for future work is the exploration of more advanced inter-car communication systems. While the current reward/penalty mechanism for shared information utility is a good start, there is potential for more nuanced and context-aware communication protocols that can further enhance the cooperative behavior of autonomous vehicles. Research into decentralized learning networks, where cars can learn from each other in real-time without relying solely on a centralized blockchain, could offer additional benefits in terms of scalability and responsiveness.

Adjusting reward mechanisms and adapting the number of goals that are given to a self-driving car could also help in shaping the reliability and stability of the self-driving cars' behavior on the road, introducing rankings, hierarchies or dynamic relationships of the various goals a self-driving car is meant to achieve (energy consumption, time to destination, respecting speed limits etc).

Finally, addressing the ethical and societal implications of autonomous vehicle deployment requires continuous dialogue among technologists, policymakers, and the public. Establishing clear guidelines for ethical AI behavior, ensuring transparency in decision-making processes, and developing robust regulatory frameworks will be essential for the responsible integration of self-driving cars into society.

Annex 1: code example

Generate.metadata.for.future.actions.based.on.raw.sensor.data?attached.to.heatmap

This code simulates the process of creating metadata for future actions based on the heatmap risk evolution, and iteratively updating these actions based on new sensor data every 50ms. The final global instructions are interpolated from all generated timelines to create a coherent set of actions for the self-driving car to follow. A final balancing algorithm arbitrates between the metadata associated to the heatmap in order to balance basic risk perception and the metadata instructions.

```python
# Function to generate metadata based on heatmap
def generate_metadata(heatmap, time_horizon=10, interval=1):
    actions = []
    for t in range(0, time_horizon, interval):
        risk_level = np.mean(heatmap[int(t*heatmap.size/time_horizon):int((t+interval)*heatmap.size/time_horizon)])
        if risk_level < 0.5:
            actions.append(f"At {t} seconds: Brake steadily until speed is 0 over {interval*5} meters")
        else:
            actions.append(f"At {t} seconds: Maintain speed for {interval} seconds")
    return actions

# Generate metadata for future actions
metadata = generate_metadata(initial_heatmap)
print(metadata)
```

```python
# Function to update sensor data and heatmap
def update_sensor_data(sensor_data, delta_t=0.1):
    # Simulate the cow moving closer
    sensor_data['cow_position'] = (sensor_data['cow_position'][0] - delta_t, sensor_data['cow_position'][1])
    return sensor_data

# Function to simulate the iterative process
def simulate_process(sensor_data, iterations=10):
    timelines = []
    for i in range(iterations):
        sensor_data = update_sensor_data(sensor_data)
        heatmap = create_risk_heatmap(sensor_data)
        metadata = generate_metadata(heatmap)
        timelines.append(metadata)
    return timelines

# Simulate the process
timelines = simulate_process(sensor_data)
for t, timeline in enumerate(timelines):
    print(f"Time {t*0.1:.1f}s: {timeline}")
```

```python
# Function to interpolate instructions from each timeline
def interpolate_instructions(timelines):
    global_instructions = []
    for timeline in timelines:
        for instruction in timeline:
            if instruction not in global_instructions:
                global_instructions.append(instruction)
    return global_instructions

# Interpolate the instructions
global_instructions = interpolate_instructions(timelines)
print(global_instructions)
```